\documentclass[manuscript,acmsmall,submission,nonacm]{acmart}
\usepackage{multirow}
\usepackage{makecell}
\usepackage{float}

\AtBeginDocument{%
  \providecommand\BibTeX{{%
    \normalfont B\kern-0.5em{\scshape i\kern-0.25em b}\kern-0.8em\TeX}}}

\setcopyright{acmcopyright}
\copyrightyear{2018}
\acmYear{2018}
\acmDOI{10.1145/1122445.1122456}

\acmJournal{TALLIP}
\acmVolume{37}
\acmNumber{4}
\acmArticle{1}
\acmMonth{8}

\begin{document}

\title{A Comparative Study of Speaker Role Identification in Air Traffic Communication Using Deep Learning Approaches}

\author{DONGYUE GUO}
\email{dongyueguo@stu.scu.edu.cn}
\orcid{0000-0003-0393-5197}
\affiliation{%
  \institution{National Key Laboratory of Fundamental Science on Synthetic Vision, College of Computer Science, Sichuan University}
  \city{Chengdu}
  \country{China}
  \postcode{610000}
}

\author{JIANWEI ZHANG}
\email{zhangjianwei@scu.edu.cn}
\author{BO YANG}
\email{boyang@scu.edu.cn}

\author{YI LIN}
\authornote{Corresponding author.}
\email{yilin@scu.edu.cn}

\affiliation{%
  \institution{College of Computer Science, Sichuan University}
  \city{Chengdu}
  \country{China}
  \postcode{610000}
}



\renewcommand{\shortauthors}{Guo and Zhang, et al.}

\begin{abstract}
Automatic spoken instruction understanding (SIU) of the controller-pilot conversations in the air traffic control (ATC) requires not only recognizing 
the words and semantics of the speech but also determining the role of the speaker. 
However, few of the published works on the automatic understanding systems in air traffic communication focus on speaker role identification (SRI).  
In this paper, we formulate the SRI task of controller-pilot communication as a binary classification problem. 
Furthermore, the text-based, speech-based, and speech and text based multi-modal methods are proposed to achieve a comprehensive comparison of the SRI task.  
To ablate the impacts of the comparative approaches, various advanced neural network architectures are 
applied to optimize the implementation of text-based and speech-based methods. 
Most importantly, a multi-modal speaker role identification network (MMSRINet) is designed to achieve the SRI task by considering both the speech and textual modality features.
To aggregate modality features, the modal fusion module is proposed to fuse and squeeze acoustic and textual representations by modal attention mechanism and self-attention pooling layer, respectively.    
Finally, the comparative approaches are validated on the ATCSpeech corpus collected from a real-world ATC environment. 
The experimental results demonstrate that all the comparative approaches are worked for the SRI task, and the proposed MMSRINet  
shows the competitive performance and robustness than the other methods on both seen and unseen data, achieving 98.56\%, and 98.08\% accuracy, respectively.
\end{abstract}



\maketitle

\section{Introduction}
Speech communication between air traffic controllers (ATCOs) and pilots is one of the most important interaction ways in air traffic control (ATC) procedures. 
Recently, there is increasing interest in introducing automatic spoken instruction understanding (SIU) techniques to empower the ATC process \cite{PardoFFRCGMSDG11, lin_spoken_2021}.  
In the past decades, it has been extensively investigated and widely applied on the 
safety detection system \cite{LinTYYZY19, LinDCWZY20}, the ATCOs training devices \cite{SmidlSTMRI19}, the ATC assistance systems \cite{9527247, 2015Assistant}. \par 

In practice, ATC speech communication can be regarded as task-oriented conversations between ATCOs and pilots. 
Thus, the SIU system of ATC is usually a pipeline that consists of several subtasks, i.e., speech activity detection (SAD), automatic speech recognition (ASR), 
text instruction understanding (TIU), and speaker role identification (SRI). 
In this pipeline, firstly, the speech segment is captured from the real-time streaming by the SAD module, and then the ASR system translates it into human-readable texts. 
Subsequently, the TIU module converts the natural texts into predefined structured instructions that are further processed by the computer. 
Finally, the computer-readable instructions and the speaker role (ATCO or pilot which was output by the SRI module) jointly provide a conversation context for other downstream applications. \par 

As can be seen from mentioned illustrations, the SRI module is a critical component of the SIU system in the field of ATC. 
However, most of the existing research of the ATC-related SIU systems focuses on the ASR and TIU techniques \cite{9174746, LIN2021107847, Zuluaga-GomezMZ20, OualilKSSHM17}, and no detailed description of the SRI task was presented. 
A instruction understanding model and ATC communication rule-based methods for the SRI tasks were studied in \cite{LinTYYZY19}, without providing SRI performance. 
To the best of our knowledge, none of the published works have reported complete approaches and results for the SRI tasks in the ATC domain.
\par 

Since the ATCO communicates with several pilots by radio in a single frequency, the role of the speaker cannot be distinguished from the communication data link. 
However, the speaker role is a kind of indispensable and important information in many ATC-related applications, such as safety detection systems, ATCO workload analysis systems.
Therefore, the inability to identify the speaker role directly from communication brings a certain challenge to the ATC-related SIU tasks. 
Fortunately, there are two kinds of data that can be served as the potential entities for the SRI tasks. 

\begin{itemize}
    \item Text: On the one hand, according to the communication rules recommended by the international civil aviation organization (ICAO), 
    the ATCOs should declare the call sign of the target aircraft before issuing the instructions, 
    while the pilots read back the instructions firstly and then reporting their call sign. 
    In general, most of the controller-pilot speech communication follow these rules, allowing the text classification to be 
    a promising technology for SRI tasks. 
    \item Speech: On the other hand, the speech can be considered as a representation of the speaker role from two aspects of signal and text. 
    a) The controller-pilot speech communication presents distinctive features depending on the equipment and environment, such as a microphone, push-to-talk (PTT), background noise, radio.  
    b) It implies the representation of its transcripts, which further provides more discriminative knowledge for the SRI task.
\end{itemize}

In this paper, we define the SRI task as a binary classification problem, i.e., all the instructions are classified into two classes: ATCO or pilot. 
Meanwhile, the SRI task is addressed by the data-driven approaches from three different inputs, i.e., text, speech, speech-text. 
To this end, the text classification, audio classification, and multi-modal classification approaches are proposed to achieve the SRI task. 
In this procedure, several popular network architectures are introduced to serve as backbone networks for each approach 
to eliminate the impact of the difference between network architectures.
The BiLSTM \cite{DBLP:conf/acl/ZhouSTQLHX16}, TextCNN \cite{DBLP:conf/emnlp/Kim14}, and Transformer \cite{DBLP:conf/nips/VaswaniSPUJGKP17} architecture are developed as the backbone network in the text-based methods, 
while x-vector \cite{snyder_x-vectors_2018}, SincNet \cite{ravanelli_speaker_2018}, and CRNN \cite{choi2017convolutional} architecture are built for the speech-based methods. 
Most importantly, a multi-modal speaker role identification network (MMSRINet) is designed to learn the distinctive 
representations from both the speech and textual modalities for the speech-text based methods. 
Specifically, a modal attention mechanism is proposed to fuse the different representations to a joint feature vector. 
In addition, the self-attention pooling layer is applied to produce the joint vector by the weighted sum operations, which further be regarded as the multi-modal embedding. 
Finally, the multi-modal embedding is further fed into the classifier to generate the final probabilities of the speaker role. 
\par 

All the proposed methods were validated on the ATCSpeech corpus \cite{yang_atcspeech_2020-1} that was collected from a real-world ATC environment. 
In addition, in order to analyze and compare the performance and robustness of the model, we evaluate the trained model in two ways:   
1) The model is validated on the test set of the ATCSpeech to evaluate the performance on the seen samples. 
2) A supplement test set called test-s is used to verify the robustness of the model on the unseen samples collected in controller-pilot communication. \par 
In summary, our contributions are listed as follows:\par 
\begin{itemize}
    \item A thorough comparison between the aforementioned deep learning based SRI techniques is investigated.  
    To the best of our knowledge, this is the first work that investigates the SRI task in the ATC domain. 
    \item The robustness and performance of the comparative models are comprehensively analyzed and discussed on the seen and unseen samples. 
    \item A multi-modal SRI network, called MMSRINet, is proposed to achieve the ATC-related SRI task by considering both the speech and textual modal knowledge, 
    which shows more competitive performance and robustness than other methods. 
\end{itemize}

This paper is organized as follows. A brief review of related works is described in Section 2. 
Section 3 presents the detail of the proposed methods and architectures. 
The experimental configurations are provided in Section 4. 
The experimental results are reported and detailly discussed in Section 5.  
Finally, this paper is concluded in Section 6.

\section{Related Work}
\subsection{Text Classification}
Text classification is a classical task in the field of natural language processing (NLP), which aims to classify a given text sequences into a certain class.  
In general, the approach can be grouped into two categories: rule-based methods and data-driven based methods. 
The rule-based approach usually requires a large number of predefined rules and is strongly dependent on domain knowledge, 
which can only be applied to limited scenarios due to poor flexibility. \par

Thanks to the development of deep learning techniques, the performance of data-driven methods has generally outperformed that of rule-based methods in recent years 
and has become the standard paradigm of text classification tasks \cite{minaee_deep_2021}. 
Zeng et al. \cite{DBLP:conf/emnlp/Kim14} utilized a convolutional neural network (CNN) \cite{726791} to achieve the sentence classification tasks which makes representative progress in the NLP domain. 
To capture the long-term dependencies, the Att-BiLSTM model \cite{DBLP:conf/acl/ZhouSTQLHX16} was built on a recurrent neural network (RNN) \cite{mikolov_recurrent_2010}. 
Currently, various improved methods based on the CNN or RNN block were proposed to achieve the text classification task, 
such as Character-level CNNs \cite{DBLP:conf/nips/ZhangZL15}, tree-based CNN \cite{DBLP:conf/acl/MouMLX0YJ16}, Tree-LSTM \cite{DBLP:conf/acl/TaiSM15}, Multi-Timescale LSTM \cite{DBLP:conf/emnlp/LiuQCWH15}. 
With the successful applications of the Transformer architecture \cite{DBLP:conf/nips/VaswaniSPUJGKP17}, 
many Transformer-based and pretrained language models were also proposed and achieved surprising performance \cite{radford_improving_2018, DevlinCLT19}. 
These methods achieved new state-of-the-art performance in text classification tasks by fine-tuning the pretrained models \cite{DBLP:conf/cncl/SunQXH19}. 

\subsection{Audio Classification}
Audio classification is widely applied in audio pattern recognition tasks, such as speaker identification \cite{snyder_x-vectors_2018,ravanelli_speaker_2018}, acoustic event detection \cite{kumar2016audio}, 
accent classification \cite{hansen2016unsupervised, lopez-moreno_automatic_2014}, audio emotion recognition \cite{jermsittiparsert2020pattern}. 
Recently, deep learning methods showed promising performance compared to traditional approaches for this task \cite{hershey_cnn_2017}. 
Enormous works have been investigated to explore different model architectures and applications for audio classification. 
Shawn Hershey et al. \cite{hershey_cnn_2017} demonstrated that the CNNs used in the image classification task, such as AlexNet \cite{krizhevsky2012imagenet}, VGG \cite{simonyan2014very}, and ResNet \cite{he_deep_2016}, 
achieved desired performance for the large-scale audio classification task. 
Meanwhile, the convolutional recurrent neural network (CRNN) was proposed and to achieve music classification \cite{choi2017convolutional}, audio event detection \cite{cakir2017convolutional}, audio tagging \cite{xu_large-scale_2018}, etc. 
In addition, in recent years, there are increasing interest in learning features from raw waveforms directly instead of handcraft features.  
Mirco Ravanelli et al. proposed the SincNet \cite{ravanelli_speaker_2018} to achieve speaker recognition which employs band-pass filters (based on the parametrized Sinc functions) in the first convolutional layer. 
Jee-weon Jung et al. proposed the RawNet \cite{jung_rawnet_2019} to improve the performance of the speaker verification from raw waveforms. 
In short, deep learning-based audio classification is still an interesting task in many applications.   

\subsection{Multi-modal Classification}
With the explosive growth of multi-modal data in the digit world, multi-modal learning is attracting increasing research interest and shows powerful performance than that of unimodal modal methods \cite{DBLP:conf/icml/NgiamKKNLN11}.  
Various modalities can be used to achieve classification tasks, including audio-video \cite{nagrani2018seeing}, audio-text \cite{mittal2020m3er}, image-text \cite{gallo_image_2018,kiela_efficient_nodate-1}, etc. 
In general, the fusion strategy of the classification task can be implemented in the following two ways: early fusion and later fusion \cite{baltrusaitis_multimodal_2019}. 
Early fusion methods fuse the multi-modal feature vectors to a joint representation that is further fed into the classifier,
while the later fusion makes a second decision on the output of two classifiers by an extra strategy. 
Since the advantages of early fusion in exploring the correlations and interactions between different modalities, in this paper, 
we introduce the early fusion methods to the SRI task. 
In an early work \cite{kiela2014learning}, direct concatenation was employed to produce multi-modal joint representations. 
In order to identify the correlations of learned multi-modal features, a structural regularization was proposed in \cite{wu2014exploring} to empower the deep neural network (DNN) based fusion layer, 
which also preserves the diversity of the different modality features. \par 
In addition to the classification task, more powerful fusion methods were successfully integrated into the ASR and NLP architectures. 
Modality attention was proposed to fuse the audio-visual features for ASR tasks in \cite{zhou_modality_2019}. 
The works of \cite{DBLP:conf/emnlp/FukuiPYRDR16} and \cite{ovalle2017gated} used compact bilinear pooling and complex gating mechanisms to obtain multi-modal representations.  

\section{Methodology}

\subsection{The Unified Framework of Unimodal SRI} \label{sec3.1}
\begin{figure*}[h]	
	\centering
    \includegraphics[width=12.7 cm]{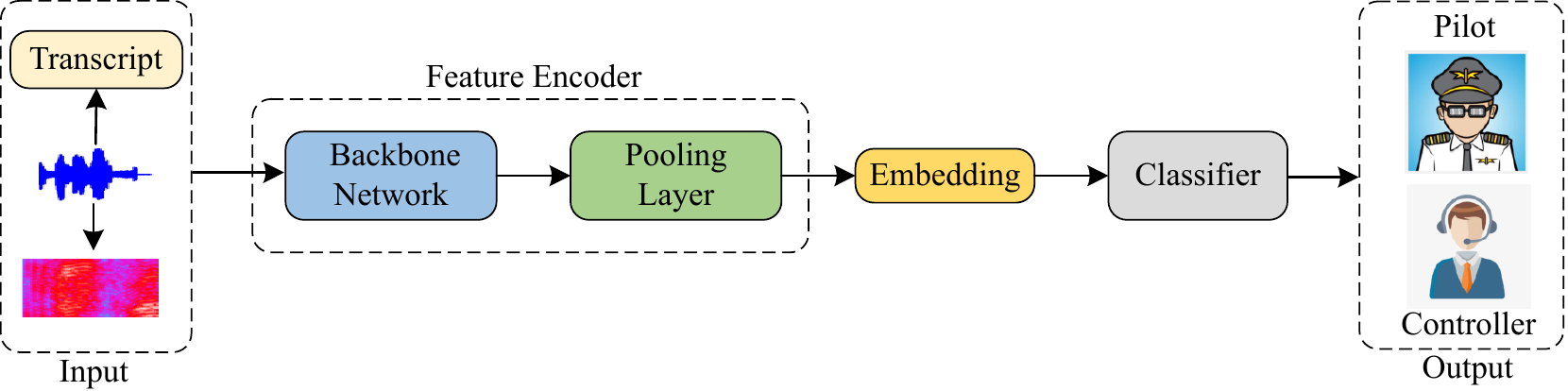}
    \caption{A unified framework of text-based and speech-based methods towards SRI tasks.\label{fig1}}
\end{figure*}

Since the SRI task is formulated as the binary classification problem, 
both the text-based and speech-based methods can be further refined as the classification task of sequential data with a variable length. 
In order to ensure the fairness of the comparison, a unified classification framework is designed for the text-based and speech-based methods. 
As illustrated in Fig. \ref{fig1}, the proposed framework includes an input module, feature encoder, and classifier. 
The detailed descriptions are shown below: \par


The input module is a conceptual component that can be compatible with both the text and speech input. 
In the text-based method, the input module is a word embedding layer to learn the representations of the transcripts. 
For the speech-based method, the input module is a speech signal preprocessor for raw waveforms, 
or a speech feature extractor for handcraft speech features. \par 

The feature encoder consists of the backbone network and a pooling layer. 
The former extracts the high-level representations from the input sequence, while the latter squeezes the feature to a fixed dimensional vector. 
Specifically, let a sequence of features $\mathbf{x} = \{x_1,x_2, ..., x_t\}$ that outputted by the input module, 
the backbone network is performed to generate the high-level feature representations $\mathbf{h} = \{h_1, h_2, ...,h_t\}$ as Eq. (\ref{eq1}), 
where $h_t \in \mathbb{R}^D$ is a D-dimensional feature vector. 
Then, as shown in Eq. (\ref{eq2}), the pooling layer squeezes the $\mathbf{h}$ to a fixed dimensional embedding vector $z \in \mathbb{R}^1$ on the temporal dimension by the pooling 
function $G$. 
Moreover, in this framework, the backbone network and the pooling layer are presented as the plugins, 
which are further adapted to inputs (text or speech) and facilitate the comparison of different algorithms. \par

\begin{equation} \label{eq1}
    \mathbf{h} = Backbone(\mathbf{x})
\end{equation}

\begin{equation} \label{eq2}
    z = G(\mathbf{h})
\end{equation}

Three fully connected (FC) layers are designed in the classifier, i.e., two hidden layers, and an output layer,  
which further transforms the embedding into the probability of the speaker role. 
The ReLU is selected as the activation function for hidden layers, while the output layer adopts Softmax activation.  
In addition, batch normalization is designed before each activation. \par 

\subsection{Text-based method}
\subsubsection{Transcription of ATC speech}
The text-based methods of the SRI task are often cascaded with the ASR system. 
In short, the transcripts of the ATC speech segments are recognized by the ASR system and serve as input for the SRI task. 
Therefore, in this case, some existing approaches can be introduced to solve this problem, such as text classification, grammar matching. \par 
The core idea of the text-based SRI approach is based on the ATC rules issued by ICAO, both of the ATCOs and the pilots are expected to speak in a rigorous stylized sentence.  
Specifically, as group A in Table \ref{tab1}, the ATCOs must indicate the call sign of the target flight before speaking details of the instructions. 
On the contrary, the pilots report their call sign after reading back the instructions in the repetition procedure. 
However, in practice, some ATC instructions break the ATC rules, which causes extra burdens for text-based approaches. 
For example, as group B in Table \ref{tab1}, the pilot instruction starts with a call sign, which deviates from the ICAO recommendations.

\begin{table}[h] 
	\centering
    \caption{The example of the ATC speech transcripts. \label{tab1}}
    \begin{tabular}{ccc}
    \toprule
    \textbf{Group}	& \textbf{Role}	& \textbf{Transcript} \\
    \midrule
    \multirow{2}*{A} & Controller	& \textit{Air China four two three seven}, climb to eight thousand one meters \\
                     & Pilot	    & Climb to eight thousand one meters, \textit{Air China four two three seven}  \\ 
    \multirow{2}*{B} & Pilot	    & \textit{Cathay two five four}, request heading two seven zero  \\
                     & Controller	& \textit{Cathay two five four}, heading two seven zero agreed \\
    \bottomrule
    \end{tabular}
\end{table}

In summary, the recommended ATC rules are the primary basis for the text-based SRI approach, which is able to achieve a preferred performance for most speech communications in the ATC environment. 
For the speech instructions that deviate from the ATC rules, the model is expected to learn discriminative features from a large amount of dataset.

\subsubsection{Network Design}
In the text-based methods, the word sequence of the ATC speech transcriptions is firstly fed into an embedding layer to generate the word representations.
In succession, the backbone network is applied to learn high-level representations that are further converted into an embedding by the pooling layer. 
Finally, the classifier (as described in Section 3.1) is designed to estimate the probabilities of the speaker role. \par 

In this work, three competitive neural architectures are selected as the backbone network to achieve text-based SRI tasks, 
i.e., bi-directional long short-term memory network (BiLSTM), CNN, and Transformer network, 
which are the well-known models for deep learning techniques. 
The detailed descriptions of the aforementioned backbone networks are summarized as follows: \par 

\begin{itemize}
    \item	\textbf{BiLSTM}: In the RNN-based architecture, two BiLSTM Layers with 512 neurons are designed to serve as the backbone network. 
    Compared to unidirectional LSTM networks, the BiLSTM networks is able to consider information from both the future and past dimensions to learn the temporal 
    dependences of the input sequence. \par
    \item	\textbf{CNN}: As referred to \cite{DBLP:conf/emnlp/Kim14}, three CNN blocks (with variable kernel size) 
    are applied to extract position-invariant representations from the embedding vector. 
    Then, the feature map is produced by cascading the output of CNN blocks. 
    The CNN block is constructed by concatenating the Conv2D layer, ReLU activation, and a Max Pooling layer. 
    The size of CNN filters is set to (3, 4, 5), corresponding to different receptive fields. \par
    \item	\textbf{Transformer}: The backbone network consists of 4 Transformer blocks, where the basic block consists of a masked multi-head attention module, 
    a layer normalization, and a position-wise feed forward layer. 
    In this work, we adopt 4 heads in the attention module, and the dimension of the feed forward layer is set to 512. \par
\end{itemize}

For the pooling layer, a self-attention based temporal pooling strategy is applied to the SRI model. 
Let the feature map $H=\{h_1, h_2,...,h_t\}$ that generated by backbone network, where $t$ is the length of the input sequence.  
The embedding $e$ is produced by a weighted sum of the feature map $H$, in which the weight is calculated by the attention mechanism. 
The inference rule can be summarized as Eq. (\ref{eq3})-(\ref{eq5}), where $H \in \mathbb{R}^{h^{b} \times t}$, $h^b$ is the dimension of the output vector, 
$\mathbf{W}$ is a trainable weight. 

\begin{equation} \label{eq3}
    H^* = tanh(H)
\end{equation}

\begin{equation} \label{eq4}
    \alpha = softmax(\mathbf{W}^T H^*)
\end{equation}

\begin{equation} \label{eq5}
    e = tanh(H\alpha^T)
\end{equation}

\subsection{Speech-based method}
\subsubsection{SRI-related features of the ATC Speech}
\begin{figure*}[tbp]
	\centering	
    \includegraphics[width=12.7 cm]{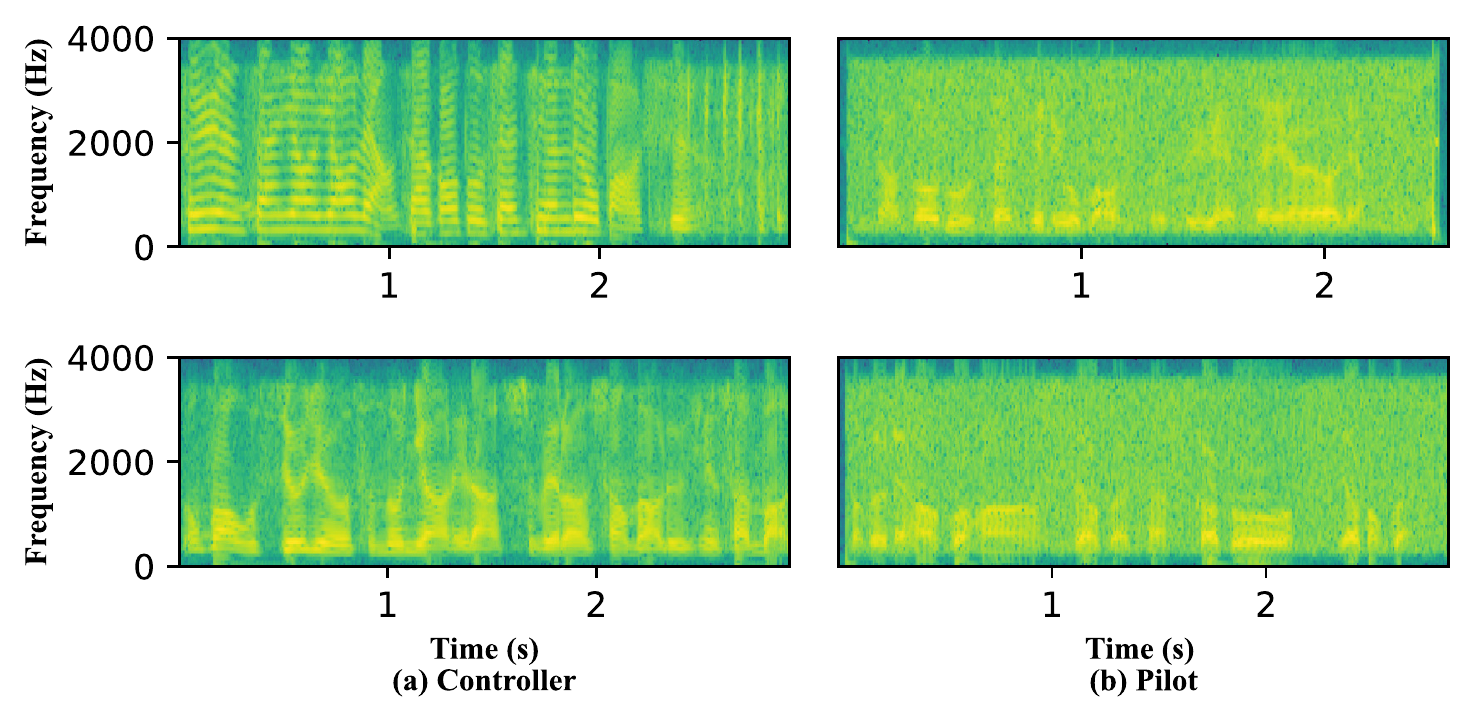}
    \caption{Selective examples of ATC speech spectrogram. \textbf{(a)} The spectrogram of the controller speech and \textbf{(b)} The spectrogram of the pilot speech. \label{fig2}}
\end{figure*}

In the ATC procedure, the speech of controller-pilot communication is transmitted through very high frequency (VHF) radiotelephony, 
in which the ATCO's speech is land-to-air whereas the pilot's speech is air-to-land. 
Thus, the special representations in the radio, microphones, and background noise (control room and aircraft cockpit) will be presented in the speech signal. 
Fig. \ref{fig2} shows the spectrogram of ATC speech, both the ATCO and the pilot speeches are collected from different speakers and control sectors. 

It is evident that the feature intensity distribute in different frequencies between ATCOs and pilots. 
For example, the frequency energy distribution of the ATCO speech is stronger than that of the pilot above 3000 Hz. 
In addition, different background noise models are presented for different speeches, specifically, the background noise distribution of the pilot speech is in a uniform manner whereas that of the ATCO is volatile.  
Moreover, the transcription implied in the speech is also the key information to support the SRI task. 
Therefore, the audio classification is expected to be a promising approach for the speech-based SRI task.

\subsubsection{Feature Encoder Design}
In the speech-based approaches, we also select three kinds of advanced neural networks of the speech-related tasks as the feature encoder, i.e., 
CRNN \cite{choi2017convolutional}, X-vector \cite{snyder_x-vectors_2018}, SincNet \cite{ravanelli_speaker_2018}. 
Since the spectrogram is a kind of detailed and visualized handcraft features and has been widely used in audio classification tasks \cite{han2016deep, zeng_spectrogram_2019}, 
the spectrogram of the raw speech serves as the input of CRNN, X-vector backbone networks. 
For the SincNet, due to its ability to process the speech signal, the raw waveform is directly fed into the model to achieve the SRI task. 
The detailed configurations of the feature encoder are described as follows:

\begin{itemize}
    \item	\textbf{CRNN}: CRNN is a popular architecture in the audio classification task. 
    In this work, the CRNN-based backbone network is implemented by referring to \cite{choi2017convolutional}. 
    The CRNN consists of 5 CNN blocks and 2-layer RNN with gated recurrent units (GRU). 
    The CNN blocks are to recognize spatial-cross patterns by filters and downsample the features by the Max-pooling operation, 
    whereas RNNs aim to summarize the learned patterns in the temporal dimension. 
    Followed the backbone network, the self-attention pooling layer is applied to generate speech embeddings that are further fed into the classifier.
    \item	\textbf{X-vector}: The X-vector system was proposed to extract DNN embeddings for speaker recognition \cite{snyder_x-vectors_2018}. 
    In this work, the front-end of the X-vector system is selected as the feature encoder, which includes 4 time-delay deep neural network (TDNN) layers and a statistics pooling layer. 
    The frame-level representations are learned by TDNN layers, and the statistic pooling layer aggregates it to the sentence-level embedding.  
    \item	\textbf{SincNet}: The SincNet is a novel and effective CNN architecture for the speaker and speech recognition tasks with raw waveforms \cite{ravanelli_speech_2018, ravanelli_speaker_2018}.
    In the SincNet block, band-pass filters are applied to replace the standard CNN filters to convolves the waveform, which shows better model convergence and performance compared to the CNN block. 
    In order to analyze the impact of the input features for the speech-based SRI tasks, instead of the handcraft features, 
    a Sinc convolutional layer is designed before the TDNN module of X-vector models to conduct the SincNet-based backbone network. 
\end{itemize}


\begin{figure*}[bp]
	\centering	
    \includegraphics[width=\linewidth]{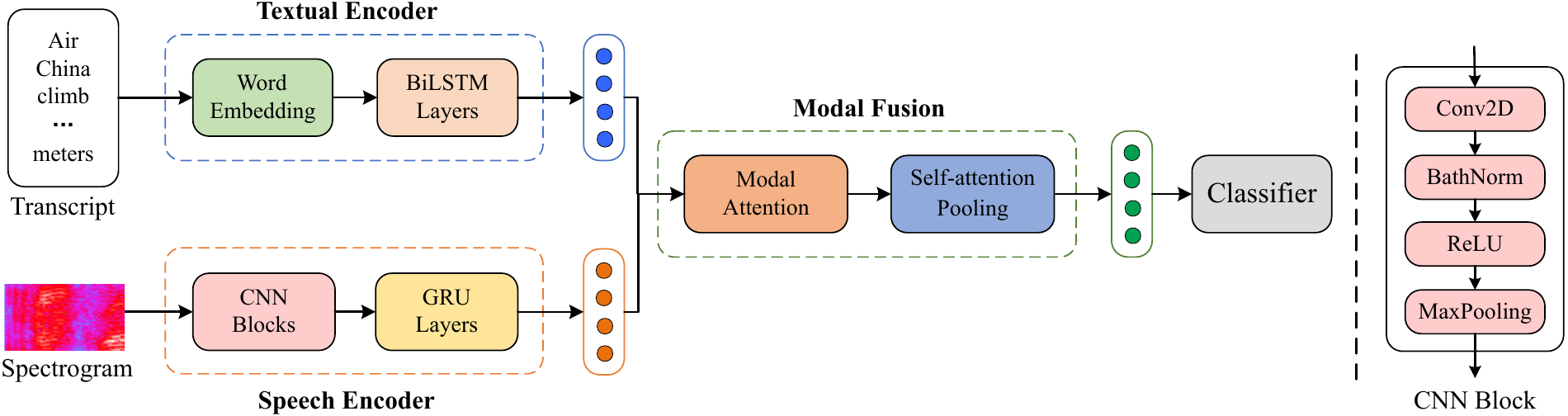}
    \caption{The architecture of the proposed MMSRINet. \label{fig3}}
\end{figure*} 


\subsection{MMSRINet} \label{sec3.4}
According to the aforementioned analysis, it is well received that the performance of text-based methods generally relies on ATC grammar, 
whereas the speech-based methods closely relate to the communication environment (equipment and background noise, etc.).
The performance of text-based methods will be significantly reduced if the speech instructions deviate from the pre-defined ATC grammar. 
Similarly, the speech-based methods will suffer poor accuracy when it works on unseen data (i.e., communication environment not be covered by the training set). 
Following this idea, we believe that the discriminative knowledge learned by the text-based method and the speech-based method are complementary and can be further used to improve the accuracy of the SRI task. 
To this end, we present a multi-modal speaker role identification network, called MMSRINet, to consider both the textual modal knowledge and the signal modal characteristics of the ATC speech.  \par

The architecture of the MMSRINet is illustrated in Fig. \ref{fig3}, consisting of the textual encoder, speech encoder, modal fusion module, and a classifier. 
Specifically, the high-dimensional textual and acoustic representations are learned by the textual encoder and speech encoder, respectively. 
Then, the representations from different modalities are aggregated and squeezed by the modal fusion module. 
Finally, the fused embedding is fed into the classifier to generate the final probability of the speaker role. \par

The textual encoder and speech encoder is implemented on the BiLSTM and CRNN, as described in Section 3.2.2 and Section 3.3.3. 
Furthermore, the composition of the CNN block also is shown in Fig. \ref{fig3}. 
The modal fusion module is constructed by the modal attention mechanism and the self-attention pooling layer, where the self-attention pooling operation is similar to that described in Section 3.2.2. 
The classifier is the same as that of in unimodal SRI framework described in Section 3.1. 
In this section, we mainly focus on fusing the representations between different modalities. 
Due to the heterogeneity of the speech and the text, aggregating these representations to a joint feature vector is still a unique computational challenge in the SRI task. 
In this work, a modal attention mechanism is designed to capture the correspondences between acoustic and textual modalities, as described below: \par 

Let the high-dimensional representations output by the speech encoder $\mathbf{h}^s=\{h_1^s,h_2^s,...,h_n^s\}$, and the textual encoder $\mathbf{h}^t=\{h_1^t,h_2^t,...,h_m^t\}$, 
where the $m,n$ are the length of the textual and speech features, respectively. 
Eq. (\ref{eq6})-(\ref{eq9}) show the fusion operation, firstly, the correlated score $e_{ij}$ between the speech feature of time step $i$ and textual feature of time step $j$ are calculated by the scoring function $Energy$ as Eq. (\ref{eq6}), 
where the $\mathbf{W}_a$ is trainable parameters. 
Secondly, the weights $\alpha_{ij}$ of the modality attention are generated by the $Softmax$ function (Eq. (\ref{eq7})). 
Thirdly, the correlated vector $c_i$ between the speech representation of time step $i$ and textual representation is computed by the weighted sum operation. 
Finally, the concatenation operator is applied to combine the correlated vector $c_i$ and speech feature $h_i^s$, 
and the final fused feature $f_i$ of time step $i$ can be produced as Eq. (\ref{eq9}), where the $\mathbf{W}_b$ is trainable parameters. 

\begin{equation} \label{eq6}
    e_{ij} = Energy(h_i^s, h_j^t) = {h_i^s}^T \mathbf{W_a} h_j^t
\end{equation}	

\begin{equation} \label{eq7}
    \alpha_{ij}=Softmax\left(e_{ij}\right)=\frac{\exp \left(e_{ij}\right)} {\sum_{k=1}^{M} \exp \left(e_{ik}\right)}
\end{equation}

\begin{equation} \label{eq8}
    c_i = \sum _{j=1}^M \alpha_{ij} h_j^t
\end{equation}

\begin{equation} \label{eq9}
    f_i = tanh(\mathbf{W_b} [c_i,h_i^s])
\end{equation}

The purpose of the modal attention mechanism is to calculate the correlations between textual representations and acoustic representations, 
so as to enhance the implicit textual information related to SRI tasks in acoustic representations, especially the position and semantic representations of the call signs. 
Further, the proposed modal attention mechanism also relieves the impacts of the communication environment, which further improves the robustness of the model. 

\section{Experimental Setup}
\subsection{The ATC Corpus}
In this work, the experiment is conducted on the ATCSpeech corpus \cite{yang_atcspeech_2020-1} that was collected from real-world ATC communications. 
The transcripts and speaker roles were manually labeled to support the approach validation. 
There are about 26.52 hours (25,765 utterances) of ATCO speech and 31.29 hours (35,895 utterances) of pilot speech. 
Some minor utterances (about 0.013 hours) without the speaker role information during the annotation were labeled $unknown$ and be excluded in our experiment. 
The sample rate of all samples in the ATCSpeech corpus are 8,000 Hz.  
In the training phase, all the proposed models are developed on the train set and tuned parameters on the dev set. \par 

To further evaluate the performance and robustness of the different methods, two test datasets are used in the evaluation phase, i.e., \textbf{test} set in ATCSpeech and a supplementary test set called \textbf{test-s}. 
The test-s dataset was collected from the Chengdu ATC center, which not be covered by the training dataset. 
The total duration of the test-s dataset is about 2 hours (1,930 utterances), and the labels of the speech are also annotated manually. 
The main purpose of introducing the test-s set is to evaluate the robustness of the model for the unseen ATC environment. 

\subsection{Comparison models}
Since the pretraining models have been shown competitive performance in language modeling and speech processing, the pretraining methods have become a new paradigm of deep learning. 
In order to present a comprehensive comparison of the deep learning based SRI models and validate the performance of the proposed unimodal SRI framework and the MMSRINet, 
the pretraining models of the speech and the text are cascaded with the classifier to develop the SRI task in the experiment. 
The detailed descriptions of the pretraining based SRI models are listed as follows. 

\begin{itemize}
    \item {\bf{Pretrained-T}}: The pretrained-T SRI model is a text-based approach that is finetuned with a pretrained BERT model \cite{DevlinCLT19} on the ATCSpeech corpus.
    In this work, the BERT model is trained on the ATCSpeech-Large corpus which contains 120,000+ transcriptions of the ATC communication utterances. 

    \item {\bf{Pretrained-S}}: The pretrained-S is a speech-based SRI model which is finetuned with a pretrained Wav2vec 2.0 model \cite{BaevskiZMA20} on the ATCSpeech corpus. 
    The Wav2vec 2.0 model is trained on an unlabeled speech corpus called RUD which includes about 2600+ hours of speech, 2.87 million utterances.
\end{itemize}

The details of the RUD and the ATCSpeech-large corpus can be found in our previous work \cite{LIN2021107847}. 

\subsection{Experimental Configurations}
To ensure the fairness of comparison between different models and methods, we used the same configuration in common blocks.  
Specifically, the dimension of the word embedding is set to 512, and the neurons of two FC layers in the classifier are set to 256 and 2, respectively. 
Both the dimension of the embedding outputted by the feature encoder (SRI framework) and modal fusion module (MMSRINet) are set to 512. \par 
 
In the experiment of speech-based methods, the spectrogram is computed by 80 linearly spaced log-filterbanks with 25ms windows and 15ms overlaps. 
In addition, for the SincNet, the raw speech waveforms are performed normalization and fed into the model directly. 
Due to the multilingual nature of the ATCspeech corpus, the Chinese character and the English word are served as the basic vocabulary tokens in the text-based methods. 
There are 1284 tokens in our vocabulary, includes 698 Chinese characters, 584 English words, and two special tokens, i.e., <PAD>, <UNK>. \par 


In this work, we construct and train all the models with the open-source deep learning framework PyTorch 1.4.0. 
The training server was configurated as follows: Ubuntu 16.04 operating system with 2*NVIDIA GeForce RTX 2080Ti GPU, Intel Xeon E5-2630 CPU, and 128 GB memory. 
The Adam optimizer with $10^{-4}$ initial learning rate is applied to all training tasks and the cross-entropy is selected as the loss function. 
An early stopping strategy is performed to terminate the training procedure by observing the SRI performance on the dev dataset. \par 

Since the samples of ATCO and pilot are imbalanced in the ATCSpeech corpus, the performance of all models is measured by the accuracy (ACC \%), 
precision, recall, F1-score (F1), and area under the receiver operating characteristic curve (AUC) on both the test set and test-s set. 
Specifically, ACC considers the ratio of samples that are correctly predicted, while AUC investigates the ability of the classifier to identify 
the ATCO utterance and pilot utterance as different thresholds are selected.
Precision, recall, F1 are commonly used metrics for binary classifiers, the pilot utterance serves as the positive class when these metrics are calculated in the experiment.

\section{Results and Discussions}
\subsection{Results of text-based Methods} \label{sec5.1}
The results of the text-based methods are reported in Table \ref{tab2}. 
For the proposed unimodal SRI framework, compared to the TextCNN and Transformer approach, BiLSTM based backbone achieves better performance on both the test and test-s sets, reaches 97.05\% and 94.97\% accuracy, respectively. 
The results benefit from the powerful temporal modeling of RNN architectures in time series data. 
The Transformer model suffers from poor accuracy and the lowest AUC in the test-s set while it obtains the higher AUC in the test set. 
It can be attributed to the Transformer model does not perform its full capacity on small-scale data sets, 
and optimizing it with enormous samples may be a promising way to achieve desired performance improvement. 
Overall, the three text-based models achieve comparable performance, and the accuracy is 96\%-97\% in the test set. 
However, the accuracy, precision, F1-score are significantly reduced on the test-s set, i.e., about 2\%. 
In general, the design of the backbone network is not the key factor that affects the model performance in the text-based SRI approaches. 

For the Pretrained-T model, it obtained better performance than the proposed unimodal models in most evaluation metrics, especially in AUC. 
It can be attributed that the pretrained model learned more discriminative features about the SRI task and achieve a stronger classification ability. 
We also recognize that the Pretrained-T model obtained comparable performance on test and test-s set and achieved higher robustness than the proposed SRI framework. 
However, these models are usually pretrained on a large amount of data and are not suitable for low-resource conditions. 

\begin{table}[H] 
    \setlength{\tabcolsep}{3.0pt}
    \caption{The results of the text-based methods. The Proc. and Reca. represent precision and recall, respectively.} \label{tab2}
    \begin{tabular}{cccccc|ccccc}
    \toprule
    \multirow{2}{*}{\textbf{Methods}} & \multicolumn{5}{c|}{\textbf{Test}} & \multicolumn{5}{c}{\textbf{Test-s}} \\ \cline{2-11} 
                                      & ACC \%           & AUC \%      & Prec. \%   & Reca. \%   & F1 \%       & ACC \%       & AUC \%      & Prec. \%   & Reca. \%     & F1 \%      \\ 
    \midrule
    \textbf{BiLSTM}                   & 97.05            & 77.86       & \bf{97.39} & 97.59      & 97.49       & 94.97        & 75.61       & \bf{95.39} & 96.58        & 95.98   \\  
    \textbf{TextCNN}                  & 96.94            & 78.70       & 96.99      & 97.90      & 97.44       & 94.61        & 79.98       & 93.84      & 97.75        & 95.75   \\
    \textbf{Transformer}              & 96.13            & 81.55       & 96.27      & 97.38      & 96.77       & 94.87        & 67.71       & 94.35      & 97.58        & 95.94  \\ 
    \textbf{Pretrained-T}             & \bf{97.13}       & \bf{97.26}  & 95.05      & \bf{97.99} & \bf{97.56}  & \bf{97.46}   & \bf{97.74}  & 94.62      & \bf{98.90}   & \bf{97.92}   \\
    \bottomrule
    \end{tabular}
\end{table}

\begin{table}[H] 
    \caption{Examples of different types of prediction error samples. In this table, \textit{akube}, \textit{scooter} are OOV tokens, 
    and the probabilities were output by the BiLSTM model.\label{tab3}} 
    \begin{tabular}{ccccc}
    \toprule
    \textbf{Type}	& \textbf{No.} & \textbf{Transcript} & \textbf{Role} & \textbf{Probability} \\
    \midrule
    \multirow{2}*{OOV} & 1 	& Hainan seven four five two, direct to \textit{akube} & Controller & 0.27 \\
                       & 2  & \makecell{\textit{Scooter} one two five, climb maintain seven \\thousand two hundred meters}  & Controller & 0.17 \\ 
    \multirow{3}*{DFG} & 3	& Descend maintain eight thousand one hundred meters              & Pilot  & 0.48 \\
                       & 4 	& Bohai seven four nine seven, confirm bemta one xray  & Pilot  & 0.39 \\
                       & 5 	& Confirm, bohai seven four nine seven  & Controller  & 0.34 \\
    \bottomrule
    \end{tabular}
\end{table}\vspace{-0.1 cm}

By analyzing the experimental results, 
two kinds of samples contribute to the performance degradation, i.e., out of vocabulary (OOV) and deviate from grammar (DFG). 
Among the misclassification samples of the unimodal SRI framework in the test-s set, the OOV samples account for about 30\%, while DFG sentences account for 70\%. 
The performance improvements of the Pretrained-T model on the test-s set are mainly benefited from the robustness of the pretrained model to the OVV tokens. 
Some selective samples with different prediction errors are listed in Table \ref{tab3}. 
The OOV tokens are mainly derived from the route waypoints (Table \ref{tab3} No.1) and airline call signs (Table \ref{tab3} No.2) that are unseen in the ATCSpeech corpus.  
Especially for airline call signs, the model is easy to be confused by the location of call signs and reports the error results in the prediction. 
The samples of DFG sentences can be divide into two types: the speech without call signs and spoken call signs break the ICAO recommendation.  
The former occurs mainly in the instructions repetition (as shown in Table \ref{tab3} No.3), while the latter tends to occur in the consultative conversation between the ATCOs and pilots (Table \ref{tab3} No.4 and No.5). 

In conclusion, processing the speech that contains OOV tokens and DFG sentence is still a huge challenge for the text-based SRI task. 
Indeed, identifying the speaker role of the DFG sentence in the ATC environment based on transcript alone and without conversation context is also a difficult task for human understanding, 
which makes a huge limitation of the text-based methods in the SRI task. 
In addition, the text-based methods are often cascaded with an ASR system in real-time SIU applications, so the final accuracy is also impacted by the ASR performance. 

\subsection{Results of Speech-based Methods and MMSRINet}
In this work, the core idea of the proposed modal attention mechanism is to utilize transcriptions to empower the implicit textual representations of the speech. 
Therefore, the MMSRINet can be considered as the variant of the speech-based methods. 
In this section, we report and discuss the results of speech-based methods and MMSRINet together. \par 

The results are reported in Table \ref{tab4}. 
For the unimodal SRI framework, the performance of speech-based methods is generally better than that of text-based methods, i.e., all of them achieve over 97\% ACC. 
However, the AUC of the speech-based methods is lower than that of text-based methods. 
As can be seen from the results, the CRNN have obtained the higher ACC in the test set, while the SincNet presented a more competitive performance in the test-s set, reaching 97.62\% and 95.54\%, respectively.  
For the Pretrained-S model, as discussed in Section \ref{sec5.1}, it also obtained better performance than that without the pretraining process and harvest the highest AUC in speech-based SRI models. 

\begin{table}[] 
    \setlength{\tabcolsep}{3.0pt}
    \caption{The results of the speech-based methods and MMSRINet.} \label{tab4}
    \begin{tabular}{cccccc|ccccc}
    \toprule
    \multirow{2}{*}{\textbf{Methods}} & \multicolumn{5}{c|}{\textbf{Test}} & \multicolumn{5}{c}{\textbf{Test-s}} \\ \cline{2-11} 
                                    & ACC \%       & AUC \%       & Prec. \%   & Reca. \%   & F1 \%         & ACC \%      & AUC \%       & Prec. \%   & Reca. \%   & F1 \%      \\ 
    \midrule
    \textbf{CRNN}                   & 97.62        & 68.44        & 98.21      & 97.80      & 98.0          & 94.30       & 69.87        & 92.91      & 98.33      & 95.54      \\  
    \textbf{X-vector}               & 97.44        & 68.89        & 96.54      & \bf{99.26} & 97.88         & 93.78       & 72.31        & 91.15      & \bf{99.58} & 95.18      \\
    \textbf{SincNet}                & 97.06        & 70.35        & 96.23      & 98.95      & 97.57         & 95.54       & 71.20        & \bf{98.94} & 93.83      & 96.32      \\ 
    \textbf{Pretrained-S}           & 98.13        & \bf{97.82}   & 98.46      & 96.57      & 98.50         & 97.56       & \bf{97.23}   & 98.41      & 95.51      & 97.98      \\ 
    \textbf{MMSRINet}               & \bf{98.56}   & 70.49        & \bf{98.83} & 98.91      & \bf{98.87}    & \bf{98.08}  & 82.03        & 97.63      & 99.16      & \bf{98.39} \\ 
    \bottomrule
    \end{tabular}
\end{table}

We further explore the incorrect samples and two key factors are regarded to affect the performance of the speech-based models. 
Firstly, compared to text-based methods, the percentage of DFG speech is significantly reduced in incorrect samples, about only 10\%. 
The noise of the speech is the primary reason to limit the model performance. 
Secondly, since the sample of the ATCOs and pilots are imbalanced in the training set, the speech-based model of the unimodal SRI framework is prone to be overfitted.  
The model prefers to predict the sample as the class most appear in the training set (pilot), which is also the reason for the lower AUC.
The AUC is greatly improved in the Pretrained-S model which might be learned a more robust speaker role related representation in the pretrained process.
\par 

The results of the MMSRINet demonstrated that utilizing both the speech and textual modality features is a feasible approach for the SRI tasks. 
The proposed MMSRINet reports the best accuracy and F1-score on both the test set and test-s set, achieving 98.56\%, 98.08\% ACC, and 98.87\%, 98.39\% F1-score, respectively. 
Actually, the acoustic features and textual representations of the ATC speech are complementary knowledge for the SRI tasks.  
In addition, by the proposed modal attention mechanism, the acoustic characteristics and ATC grammars of the speech can be properly considered in the proposed MMSRINet. \par 

In summary, the following conclusions can be obtained from the experimental results:
\begin{enumerate}
\item	The text-based methods usually rely on the ATC grammars and are suitable for the general conversation contexts. 
        The OOV issue is a challenge when the model is migrated to an unseen ATC environment.
\item	Speech-based methods obtain better ACC in the test set than text-based methods. 
        The representation difference in the acoustic features between ATCOs and pilots is regarded as a piece of key knowledge for the SRI tasks.
        Since acoustic features are prone to be changed and distorted in unseen or noisy channels, the performance of the speech-based methods will be limited by the real-time communication environment. 
        
\item   Pretraining based models can learn more useful SRI-related features on pretraining process and improve the robustness and classification ability of the SRI models on unseen data.  
        But it requires large amounts of data to fit the pretrained model in pretraining process, not suitable for the low-resource conditions.  

\item	The experimental result demonstrates that the multi-modal approach is a competitive solution for the SRI task. 
        It learns the complementarity knowledge from both the acoustic features and textual grammar of the ATC speech, providing more discriminative information for the classifier. 
        Thanks to the multi-modal inputs, the model presented better performance and robustness which is beneficial for unseen datasets.   
\end{enumerate}

\subsection{Ablation Study}
In this section, we design kinds of ablation experiments to verify the effectiveness of the proposed self-attention pooling strategy and modal attention mechanism in the MMSRINet. 
All experiments are conducted in the ATCSpeech corpus and evaluated on the test set. 

\subsubsection{Self-attention pooling vs. other pooling strategy}
To validate the effectiveness of the proposed self-attention pooling strategy, the sum and average operation on the temporal dimension are applied to replace the self-attention pooling in the MMSRINet. 
The results are presented in Table \ref{tab5}, it can be seen from the experimental results that the sum operation is slightly inferior to the self-attention pooling strategy and obtained the 98.37 \% accuracy and 98.64 \% F1-score. 
The performance of the average pooling operation only achieve 97.62 \% accuracy in the test set. 
This is because the average pooling strategy assumes that the features of each time step make the same contributions of the SRI task.  
In contrast, the proposed self-attention pooling strategy generates a weight coefficient at each time step, which is conducive to highlighting SRI-related features and improving the performance of the SRI task.

\begin{table}[h]
    \caption{The results of the different pooling strategy in MMSRINet.} \label{tab5} 
    \begin{tabular}{cccccc}
    \toprule
    \textbf{Pooling strategy} & \textbf{ACC\%} & \textbf{AUC\%} & \textbf{Precision \%} & \textbf{Recall \%} & \textbf{F1 \%} \\ 
    \midrule
    \textbf{Average}          & 97.62          & 64.59          & 96.64                 & 99.47              & 98.04          \\
    \textbf{Sum}              & 98.37          & 69.21          & 98.43                 & 98.84              & 98.64          \\
    \textbf{Self-attention} (Ours)   & 98.56          & 70.49          & 98.83                 & 98.91              & 98.87          \\ 
    \bottomrule
    \end{tabular}
\end{table}

\subsubsection{Modal attention vs. concatenation}
As described in Section \ref{sec3.4}, the motivation of the proposed modal attention mechanism is to utilize transcriptions to empower the implicit textual representations of the speech. 
To validate the effectiveness of the proposed modal attention mechanism, the concatenate operation is applied to alternate the modal attention and generated the joint representation vector. 

The results are shown in Table \ref{tab6}, the performance of proposed modal attention is superior to that of the concatenate operation in most evaluation metrics. 
By analyzing the misclassification samples, we found that the performance will be reduced using concatenate operation when the transcription can not provide discriminative features for the SRI task
(Such as the DFG sentence No.3 and No.5 in Table \ref{tab3}.). 
Note that it also harvests better performance than unimodal methods when concatenating the speech and textual representation directly. 
Thus, it is an effective technique to achieve the high accuracy SRI tasks by considering multi-modal features that further support our motivation.

\begin{table}[h]
    \caption{The results of concatenation operation vs. the proposed modal attention module.} \label{tab6}
    \begin{tabular}{cccccc}
    \toprule
    \textbf{Fusion methods} & \textbf{ACC\%} & \textbf{AUC\%} & \textbf{Precision \%} & \textbf{Recall \%} & \textbf{F1 \%} \\
    \midrule
    \textbf{Concatenation}  & 98.25          & 65.50          & 97.83                 & 99.26              & 98.54         \\
    \textbf{Modal attention} (Ours)   & 98.56          & 70.49          & 98.83                 & 98.91              & 98.87          \\ 
    \bottomrule
    \end{tabular}
\end{table}

In short, the above ablation experiments demonstrate that the proposed self-attention pooling strategy and modal attention module are helpful to improve the performance of the SRI models. 

\section{Conclusions}
In this paper, we presented a comprehensively comparative study for the SRI tasks using deep learning approaches in the ATC domain. 
Three kinds of methods with different inputs were investigated to solve the problems of the SRI tasks, i.e., text-based methods, speech-based methods, and multi-modal methods. 
Firstly, we formulated the SRI task as the binary classification problem, and further refine the above methods as text classification, speech classification, and multi-modal classification task. 
Secondly, the efficacy of the above methods is confirmed by theoretical and experimental demonstrations. 
Finally, the experiments demonstrated that the proposed MMSRINet is a competitive approach that achieves the best performance and robustness in the seen and unseen ATC environments. \par 

In the future, we plan to explore more efficient and effective approaches for the SRI tasks. 
In addition, the fusion and application of the multi-modal data in the ATC environment would be also an interesting research topic.

\begin{acks}
This work was supported by the National Natural Science Foundation of China under Grants 62001315 and U20A20161, 
the Open Fund of Key Laboratory of Flight Techniques and Flight Safety, Civil Aviation Administration of China (CAAC) under Grant No. FZ2021KF04, 
and Fundamental Research Funds for the Central Universities under Grant No. 2021SCU12050. 
The authors would like to thank all contributors to the ATCSpeech corpus. 
\end{acks}

\bibliographystyle{ACM-Reference-Format}
\bibliography{mybibtex}

\end{document}